\documentclass[useAMS,usenatbib]{mn2e}

\usepackage{graphicx}
\usepackage{amsmath}
\usepackage{siunitx}
\usepackage{mathrsfs}
\usepackage[]{txfonts}
\usepackage{subfigure}
\usepackage{epsfig}
\usepackage{booktabs}
\usepackage{url}
\def\simless{\mathbin{\lower 3pt\hbox
{$\rlap{\raise 5pt\hbox{$\char'074$}}\mathchar"7218$}}}   %< or of order
\def\simmore{\mathbin{\lower 3pt\hbox
{$\rlap{\raise 5pt\hbox{$\char'076$}}\mathchar"7218$}}}   %> or of order
\newcommand{\be}{\begin{equation}}
\newcommand{\ee}{\end{equation}}
\newcommand       \bea          {\begin{eqnarray}}
\newcommand       \eea          {\end{eqnarray}}
\newcommand       \apj          {ApJ}
\newcommand       \apjl         {ApJL}
\newcommand       \aap          {A\&A}
\newcommand       \nat          {Nature}
\newcommand       \apss        {APSS}
\newcommand       \mnras        {MNRAS}

\newcommand       \prd      {Phys.~Rev.~D.~}

\newcommand       \araa      {ARA\&A}

\newcommand      \apjs {ApJ Supplements}

\newcommand \physrep {Physics Reports}

\def\simlt{\mathrel{\hbox{\rlap{\hbox{\lower4pt\hbox{$\sim$}}}\hbox{$<$}}}}
\def\simgt{\mathrel{\hbox{\rlap{\hbox{\lower4pt\hbox{$\sim$}}}\hbox{$>$}}}}
\def\lesssim{\mathrel{\hbox{\rlap{\hbox{\lower4pt\hbox{$\sim$}}}\hbox{$<$}}}}

\def\gtrsim{\mathrel{\hbox{\rlap{\hbox{\lower4pt\hbox{$\sim$}}}\hbox{$>$}}}}

\title[]{Pair Fireball Precursors of Neutron Star Mergers}

\author[]{Brian~D.~Metzger$\thanks{E-mail: bmetzger@phys.columbia.edu}$, Charles Zivancev\\
Columbia Astrophysics Laboratory, Columbia University, New York, NY, 10027, USA\\}

\begin{document}
\date{Received / Accepted}
\pagerange{\pageref{firstpage}--\pageref{lastpage}} \pubyear{2012}

\maketitle

\label{firstpage}

\begin{abstract}
If at least one neutron star (NS) is magnetized in a binary NS merger, then the orbital motion of the conducting companion during the final inspiral induces a strong voltage and current along the magnetic field lines connecting the NSs.  If a modest fraction $\eta$ of the extracted electromagnetic power extracted accelerates relativistic particles, the resulting gamma-ray emission a compact volume will result in the formation of an electron-positron pair fireball.  Applying a steady-state pair wind model, we quantify the detectability of the precursor fireball with gamma-ray satellites.  For $\eta \sim 1$ the gamma-ray detection horizon of $D_{\rm max} \approx 10(B_{\rm d}/10^{14}$ G)$^{3/4}$ is much closer than the Advanced LIGO/Virgo horizon of 200 Mpc, unless the NS surface magnetic field strength is very large, $B_{\rm d} \gtrsim 10^{15}$ G.  Given the quasi-isotropic nature of the emission, mergers with weaker NS fields could contribute a nearby population of short gamma-ray bursts.  Power not dissipated close to the binary is carried to infinity along the open field lines by a large scale Poynting flux.  Reconnection within this outflow, well outside of the pair photosphere, provides a potential site for non-thermal emission, such as a coherent millisecond radio burst.  

\end{abstract} 
  

\section{Introduction} 
\label{sec:intro}

Mergers of compact object binaries containing neutron stars (NS) and stellar mass black holes (BH) are primary sources for the detection of gravitational waves (GW) by the ground-based interferometers Advanced LIGO (\citealt{Harry+10}), Advanced Virgo \citep{Degallaix+13}, and the Japanese cryogenic detector KAGRA \citep{Somiya12}.  Advanced LIGO recently announced the discovery of a BH-BH merger event (\citealt{Abbott+16}), providing the first direct detection of GWs and ushering in a new era of `GW astronomy'.

A primary goal of GW astrophysics is the discovery of an electromagnetic (EM) counterpart coincident with the GW signal (see \citealt{Bartos+13}, \citealt{Fernandez&Metzger16} for reviews).  After reaching design sensitivity, the current generation of GW detectors will be capable of detecting NS-NS mergers out to an average distance $\approx 200$~Mpc, and NS-BH mergers to a distance $\approx 2-3$ times larger \citep{Abadie+10}.  The network of interferometers can narrow down the sky position of a source, primarily through triangulation based on the GW arrival time.  Depending on the signal to noise ratio, however, the uncertainty in localization can still be tens or hundreds of square degrees (\citealt{Fairhurst11}; \citealt{Nissanke+13}; \citealt{Singer&Price16}).  This greatly exceeds the field of view of most follow-up telescopes, especially at optical and soft X-ray wavelengths (e.g, \citealt{Coughlin&Stubbs16}).  

Promising EM counterparts of NS-NS and NS-BH mergers include optical/infrared transients powered by the radioactive decay of freshly synthesized neutron-rich nuclei (\citealt{Li&Paczynski98,Metzger+10}) and short duration gamma-ray bursts (GRB) (\citealt{Berger14}, for a review).  GRBs are believed to result from the dissipation of bulk kinetic or magnetic energy within a collimated relativistic outflow.  The latter may accompany the rapid accretion of a remnant debris disk onto the BH following a NS-NS or NS-BH merger event (e.g.,~\citealt{Rezzolla+11}, \citealt{Ruiz+16}).  

In some ways, GRBs provide an ideal EM counterpart.  Their limited durations allow for an unambiguous time association with the GW chirp (\citealt{Abadie+12}), while extant satellites, including {\it Swift}, {\it Fermi}, and {\it Integral}, provide nearly continuous coverage of the hard X-ray and gamma-ray sky.  Their major drawback, however, is that the relativistic jet is believed to be confined to a narrow solid angle; due to relativistic beaming, only observers within the opening angle of the jet observe the gamma-ray emission and its synchrotron X-ray and optical afterglow.  Although the opening angles of short GRB jets are measured or constrained in only a handful of cases (e.g.~\citealt{Fong+14}), a comparison between the observed local volumetric rate of short GRBs with measured redshifts (\citealt{Wanderman&Piran15}) and the anticipated NS-NS merger rate (\citealt{Kim+15}) indicates that only a few percent of GW-detected mergers will be accompanied by observable GRBs (\citealt{Metzger&Berger12}).  In this case, tens or hundreds of GW events would need to be discovered before one with an associated GRB.  This motivates considering more isotropic sources of high energy emission, which still exploit the advantages of all-sky gamma-ray monitors, but could be detected even if the GRB itself is beamed away from our line of site or if the formation of a post-merger relativistic jet is not always successful.  The latter is particularly likely in the case of a NS-BH merger if the BH mass is sufficiently large that the NS is swallowed whole before being tidally disrupted (e.g.~\citealt{McWilliams&Levin11}).  

Compared to the post-merger phase, less theoretical work been dedicated to the EM emission during the late inspiral phase prior to coalescence.  If at least one NS is magnetized, then the orbital motion of the conducting companion NS or BH through its dipole magnetic field induces a strong voltage and current along the magnetic field lines connecting the two objects (\citealt{Lipunov&Panchenko96,Vietri96,Hansen&Lyutikov01,McWilliams&Levin11,Piro12,DOrazio&Levin13,Palenzuela+13a,Ponce+14,DOrazio+16}), in analogy with the uni-polar inductor model for the Jupiter-Io system (\citealt{Goldreich&LyndenBell69}).  This voltage accelerates charged particles, potentially powering electromagnetic emission that increases in strength as the orbital velocity increases and the binary separation decreases approaching merger.  

As shown by \citet{Lai12} and \citet{Palenzuela+13a}, the magnetic field geometry of the merging binary system is complex and time-dependent, with the magnetic field lines connecting the neutron stars being periodically torn open to infinity.  Averaged over many orbits, a fraction of the power dissipated in the circuit is dissipated as ``heat" (charged particle acceleration) in the magnetosphere, while the remainder is carried to large distances by a Poynting flux along the open field lines.  In addition to magnetosphere interaction, tidal resonant excitation of modes in the NS crust of provides an additional way to tap into the orbital energy of the merging binary \citep{Tsang+12,Tsang13}.  If such modes shatter the crust, then $\sim 10^{46}-10^{47}$ erg may be released seconds prior to merger, some fraction of which will also couple to the magnetosphere in the form of outwardly-propagating Alfven waves.  

Particle acceleration in the magnetosphere gives rise to non-thermal (e.g. synchrotron) emission.  However, this radiation is unlikely to escape to the distant observer, because such a high density of $\gtrsim$ MeV photons within such a compact volume will result in copious electron-positron pair production via $\gamma-\gamma$ annihilation (\citealt{Usov92}).  For this reason we hypothesize that the final stages of the inspiral are instead best characterized as a baryon-free ``pair fireball", similar to that developed in the context of GRBs (\citealt{Paczynski86,Goodman86,Shemi&Piran90}).  This results in an enormous simplification, because the  fireball emission is predicted to be quasi-thermal, with a temperature that depends only on the total power output and physical scale of the system and is independent of the messy details of the particle acceleration and radiative processes close to the binary.  The signal emerges at hard X-ray/gamma-ray frequencies over a timescale of milliseconds, potentially resembling a sub-luminous short GRB, albeit one which is quasi-isotropic compared to the geometrically-beamed, ultra-relativistic jet which may accompany the post-merger phase.

%This Letter is organized as follows.  In $\S\ref{sec:model}$ we describe a simple model for the energy released by the inspiral and the resulting thermal emission.  In $\S\ref{sec:GRB}$ we quantify the detectability of the signal with present gamma-ray satellites.  In $\S\ref{sec:conclusions}$ we summarize our conclusions.

\section{Precursor Emission Model}
\label{sec:model}

\begin{figure}
\includegraphics[width=0.5\textwidth]{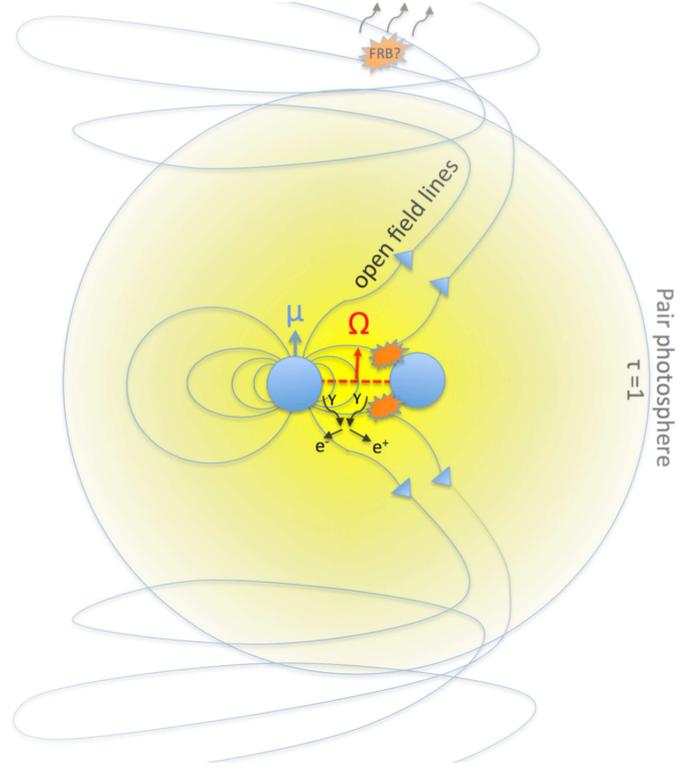}
\centering
\caption{Schematic diagram of the pair fireball created during the final stages of the NS-NS binary inspiral.  The orbital motion of one NS through the magnetosphere of the companion drives a current along the polar magnetic field lines connecting the stars.  This current generates a toroidal magnetic field which is comparable in strength to the poloidal field, the pressure-driven inflation of may cause the field lines to periodically tear open in reconnection events (\citealt{Lai12}).  Relativistic particles accelerated by unshielded electric fields, or by magnetic reconnection events near the companion, radiate energetic gamma-rays $\gtrsim 1$ MeV, which via $\gamma-\gamma$ annihilation produce copious electron-positron pairs.  A fraction of the electromagnetic energy released by the circuit thus emerges as a quasi-spherical thermal pair fireball, with the remainder being carried outwards in a large-scale Poynting flux along the magnetic field lines.  Dissipation and particle acceleration within the magnetized region well above the pair photosphere could, speculatively, give rise to a millisecond coherent radio pulse, similar to the observed fast radio bursts.  }
\label{fig:cartoon}
\end{figure}

The semi-major axis $a$ of a circular binary, comprised of two equal masses $M$, decreases due to gravitational radiation according to
\begin{equation}
\label{eq:adot}
         \frac{1}{a}\frac{da}{dt} = -\frac{128}{5}\frac{G^3M^3}{c^5 a^{4}},
\end{equation}
where we have neglected post-Newtonian corrections, which become relevant only near the very end of the inspiral.  The solution of equation (\ref{eq:adot}) is given by
\begin{equation}
\label{binary_sep}
	a^4=a_0^4(1-t/t_{\rm m,0}),
\end{equation}
where $a(t=0) \equiv a_0$ is the initial separation, $t_{\rm m,0} \equiv t_{\rm m}(a_0)$, and
\begin{equation}
\label{eq:tm}
	t_{\rm m} = \frac{5}{512}\frac{c^5a^4}{G^3M^3} = 3.0\left(\frac{a}{30{\,\rm km}}\right)^{4}\,{\rm ms}
\end{equation}
is the merger time at which the separation formally reaches zero, where in the second equality and hereafter we assume that $M = 1.4 M_{\odot}$.  In practice, the closest binary separation of interest is the point of tidal disruption or physical collision, $a_{\rm min} \gtrsim 2-3 R \approx 30$ km, where $R \approx 10$ km is the NS radius.   

\subsection{Magnetospheric Dissipation}

Following \citet{Piro12}, the orbital velocity $\vec{v}$ of the conducting companion in the magnetosphere of the primary NS induces a motional electric field $\vec{E} = -(\vec{v}\times \vec{B})/c$ across the companion, producing a potential difference of magnitude $\Phi \approx 2 R|\vec{E}| \sim 2 Ra \Omega B/c$, where $B \simeq B_{\rm d}(a/R)^{-3}$ is the dipole magnetic field of the primary NS at the location of the secondary $B_{\rm d}$ is the surface magnetic field of the primary of radius $R$, and $\Omega = (2GM/a^{3})^{1/2}$ is the orbital frequency.  Here we neglect NS spin, which is usually small compared to the orbital frequency near the end of the inspiral.  The voltage $\Phi$ drives a current along the circuit connecting the two stars.  If the resistance of the circuit is dominated by the magnetosphere instead, with a value equal to the impedance of free-space, then the resulting ohmic dissipation rate is given by
\be
\dot{E} = \frac{\Phi^{2}c}{4\pi}  \approx \frac{2}{\pi}\frac{GMB_{\rm d}^{2}R^{8}}{c a^{7}} \approx 1.8\times 10^{46}\left(\frac{B_{\rm d}}{10^{14}\,{\rm G}}\right)^{2} \left(\frac{a}{30{\,\rm km}}\right)^{-7}\,{\rm erg\,s^{-1}}.
\label{eq:edot}
\ee
For the non-recycled pulsar produced in conventional field channels of NS-NS binary formation, we expect $B_{\rm d} \sim 10^{12}-10^{13}$ G (\citealt{Bhattacharya&vandenHuevel91}).  However, it is difficult to exclude the possibility of stronger magnetic fields in the small fraction of youngest systems.  Although we have assumed a dipole field geometry, note that during the final stages of the inspiral when $a \sim$ few $R$, higher order moments of the magnetic field may come to dominate the dipole, increasing the strength of the signal.

\citet{Lai12} show that $\dot{E}$ is capped by the maximum current which can be sustained before the toroidal magnetic field induced by the current becomes comparable to that of the original poloidal dipole field.  This is because the magnetic pressure of the toroidal field causes the magnetosphere to inflate, tearing open the field lines and disconnecting the circuit (e.g.,~\citealt{Uzdensky04}).  This maximum power is given by (\citealt{Lai12})
\begin{eqnarray}
\dot{E}_{\rm max} &\simeq& \zeta_{\phi}\frac{a \Omega}{c}\frac{B_{\rm d}^{2}R^{8}c}{2 a^{6}} \approx 7\times 10^{46}\left(\frac{B_{\rm d}}{10^{14}{\rm G}}\right)^{2}\left(\frac{a}{30\,{\rm km}}\right)^{-13/2}\,{\rm erg\,s^{-1}}, \nonumber \\
&\approx& 4\times 10^{47}\left(\frac{B_{\rm d}}{10^{14}{\rm G}}\right)^{2}\left(\frac{t_{\rm m}}{\rm ms}\right)^{-13/8}\,{\rm erg\,s^{-1}}, 
\label{eq:edotmax}
\end{eqnarray}
where $\zeta_{\phi}$ is a parameter of order unity which we take equal to 1.  In practice $\dot{E} \sim \dot{E}_{\rm max}$ across most of the relevant parameter space, so we take $\dot{E} = \dot{E}_{\rm max}$ in estimates hereafter.     

We assume that a fraction $\eta <1$ of $\dot{E}$ is used to accelerate particles locally within the magnetosphere and that the remainder emerges as a Poynting flux along the open magnetic field lines.  This locally dissipated power $\sim \eta \dot{E}$ will be deposited in a region of characteristic size $L$, where $R \lesssim L \lesssim a$ defines the relevant range of physical scales in the magnetosphere.  The details of the particle acceleration process are uncertain, and may involve both unshielded electric fields which develop at localized regions within the magnetosphere, as well as periodic magnetic reconnection events near the companion.  However, the net effect is likely to be similar: the acceleration of non-thermal electron-positron pairs, for which the synchrotron cooling time is very short due to the presence of a strong magnetic field (e.g.~\citealt{Wang+16}).  Balancing stochastic acceleration with synchrotron cooling results in a universal value of the maximum synchrotron frequency $(h\nu)_{\rm max} \approx 160$ MeV (\citealt{Guilbert+83}).  Synchrotron emission extending to higher frequencies beyond the synchrotron burn-off limit is possible if the emitting particles are accelerated by a strong unshielded electric field in a region such as a current sheet or null point where the magnetic field is comparatively weak (e.g.~\citealt{Cerutti+13}, \citealt{Lyutikov+16}).  

A compact source of photons of energy $\gg 2 m_e c^{2} \approx$ 1 MeV will result in copious pair production via $\gamma-\gamma$ annihilation.  If the radiation is released within a volume $V \lesssim L^{3}$, then the energy density of the radiation is given by $u_{\rm rad} \sim f_{\rm thr}\eta\dot{E}/(4\pi L^{2}c),$ where $f_{\rm thr}$ is the fraction of the luminosity above the pair creation threshold.  The optical depth of a pair producing photon is commonly quantified by the dimensionless compactness parameter,
\be
\ell = \frac{u_{\rm rad} \sigma_T L}{m_e c^{2}} \sim 4\times 10^{10} f_{\rm thr}\eta \left(\frac{L}{a}\right)^{-1}\left(\frac{B_{\rm d}}{10^{14}\,{\rm G}}\right)^{2}\left(\frac{a}{30\,{\rm km}}\right)^{-15/2}
\ee  
For $\eta, f_{\rm thr} \sim 1$, we have $\ell \gg 1$ for $a \lesssim 1000$ km and hence the dissipation region will be opaque to pair-producing photons, and the dissipated energy $\eta\dot{E}$ will be thermalized, during the final stages of the inspiral.   A similar picture of magnetosphere pair fireball formation was described by \citet{Usov92} in the context of millisecond magnetars.  

\subsection{Pair Wind Model}

In order to study the thermal component of the precursor emission, we adopt the steady-state spherically-symmetric optically-thick pair wind model of \citet{Paczynski86}, originally developed for gamma-ray bursts.  For simplicity, we assume that the power $\eta\dot{E}(t)$ at time $t$ is deposited over a region of size $r_0 \sim a(t)$.  Hereafter we adopt $\eta = 1$ for simplicity, but in \S\ref{sec:conclusions} we speculate about the consequence of the remaining fraction $1 - \eta \sim \mathcal{O}(1)$ of $\dot{E}$ which emerges as a large-scale Poynting flux.  Although a spherical outflow is a crude approximation, if thermal pressure dominates the flow should evolve to be become quasi-spherical on scales characterizing the photosphere radius, which is typically several times larger than the binary separation near the end of the inspiral when most of the power emerges (see eq.~\ref{eq:rphoto} below).

%On larger physical scales, this magnetized outflow may produce an additional component of non-thermal emission.    

\begin{figure}
\includegraphics[width=0.5\textwidth]{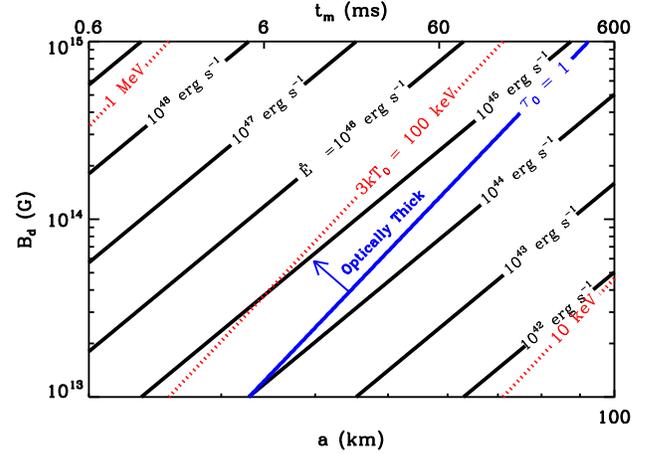}
\centering
\caption{Contours of the power dissipated in the magnetosphere, $\dot{E}$ (black lines) in the space of surface magnetic field strength, $B_{\rm d}$ and binary separation, $a$ (or, equivalently, time until merger, $t_{\rm m}$).  Red dashed lines show the characteristic photon energy of the emission $\sim 3 k T_0$ (eq.~\ref{eq:T0}).  The optically thick pair fireball model we have employed is valid only to the left of the solid blue line, which delineates the region of parameter space where the Thomson optical depth of the pairs exceeds unity near the base of the outflow ($\tau_0 = 1$; eq.~\ref{eq:tau0}).   
}
\label{fig:contour}
\end{figure}

The temperature at the base of the radiation-dominated pair flow is given by \citep{Paczynski86}
\begin{eqnarray}
\label{eq:T0}
	T_0 &=& \left(\frac{3\dot{E}}{64\pi\sigma r_0^2}\right)^{1/4} \approx 1.2\times 10^{9} \left(\frac{B_{\rm d}}{10^{14}\,{\rm G}}\right)^{1/2}\left(\frac{r_0}{a}\right)^{-1/2}\left(\frac{a}{\rm 30\,km}\right)^{-17/8}\,\,{\rm K} \nonumber \\
&\approx& 2.2\times 10^{9} \left(\frac{B_{\rm d}}{10^{14}\,{\rm G}}\right)^{1/2}\left(\frac{r_0}{a}\right)^{-1/2}\left(\frac{t_{\rm m}}{\rm ms}\right)^{-17/32}\,\,{\rm K}
\end{eqnarray}
In the limit $kT \ll m_e c^{2} \sim 5\times 10^{9}$ K of interest, the number density of electron-positron pairs in thermal equilibrium is given by
\begin{equation}
\label{eq:npair}
	n_{\pm} \simeq \SI{4.4e30}{{cm}^{-3}}\left(\frac{kT}{m_e c^2}\right)^{1.5}\exp\left[-\frac{m_e c^{2}}{kT}\right]
\end{equation}
The Thomson optical depth of the pairs near the base of the photosphere may be estimated as
\begin{equation}
\label{eq:tau0}
	\tau_0 = n_\pm(T_0) \sigma_T r_0.
\end{equation}
Fig.~\ref{fig:contour} shows the $\tau_0 = 1$ line in the space of $a-B_{\rm d}$.  Given the exponentially sensitive temperature dependence of $n_{\pm}$, in practice $\tau = 1$ is satisfied for $T \approx 0.05 m_e c^{2}/k \approx 3\times 10^{8}$ K, nearly independent of $a$.  From equation (\ref{eq:T0}), the base of the outflow is optically thick to pairs for binary separations and merger times,
\be
a \lesssim  57\,\,{\rm km}\,\,\left(\frac{r_0}{a}\right)^{-4/17}\left(\frac{B_{\rm d}}{10^{14}\,{\rm G}}\right)^{4/17},t_{\rm m} \lesssim 39\left(\frac{r_0}{a}\right)^{-16/17}\left(\frac{B_{\rm d}}{10^{14}\,{\rm G}}\right)^{16/17}\,{\rm ms}
\ee  
Thus, for $B_{\rm d} \sim 10^{13}-10^{15}$ G, the optically thick assumption is valid from $\sim 1-1000$ ms prior to the merger.  
Radiation pressure accelerates the optically-thick plasma to relativistic velocities, with the bulk Lorentz factor increasingly approximately linearly with radius $r$ from the launching point, $\Gamma(r) \simeq r/r_0$ (\citealt{Paczynski86}).  Photons escape near the photosphere radius, 
\be
 r_{\rm ph} = r_0\frac{T_{0}}{T_{\rm \tau = 1}} \approx 120{\rm\,\, km}\,\,\left(\frac{r_0}{a}\right)^{1/2}\left(\frac{B_{\rm d}}{10^{14}\,{\rm G}}\right)^{1/2}\left(\frac{a}{\rm 30\,km}\right)^{-9/8},
\label{eq:rphoto}
\ee
where the final asymptotic Lorentz factor is achieved,
\be
\Gamma_{\rm f} \simeq \frac{r_{\rm ph}}{r_{0}} \approx 4 \left(\frac{r_0}{a}\right)^{-1/2}\left(\frac{B_{\rm d}}{10^{14}\,{\rm G}}\right)^{1/2}\left(\frac{a}{\rm 30\,km}\right)^{-17/8}.
\ee
The outflow thus achieves mildly relativistic values of $\Gamma_{\rm f} \sim 2-10$ near the final stages of the inspiral for $B_{\rm d} \sim 10^{13}-10^{15}$ G.  

As internal energy is converted into kinetic energy, the temperature of the flow decreases with radius $T(r) = T_0r_0/r$.  Remarkably, however, the {\it observed} radiation temperature $T_{\rm obs} = T\Gamma = T_0$, accounting for the relativistic Doppler shift, equals the temperature at the base of the outflow (\citealt{Paczynski86,Goodman86}).

In the above we have treated the outflow as a steady-state wind at each binary separation $a$.  This is justified because the time for the outflow to reach the photosphere radius, $t_{\rm ph} \sim r_{\rm ph}/c$, is short compared to the GW inspiral time over which the power input $\dot{E}$ is changing (eq.~\ref{eq:tm}),
\be
\frac{t_{\rm ph}}{t_{\rm m}} \approx 0.13\left(\frac{r_0}{a}\right)^{1/2}\left(\frac{B_{\rm d}}{10^{14}\,{\rm G}}\right)^{1/2}\left(\frac{a}{\rm 30\,km}\right)^{-41/8},
\ee
at all stages of the inspiral ($a \gtrsim 30$ km) for $B_{\rm d} \lesssim 10^{15}$ G.

The spectrum of radiation is quasi-thermal, with a specific flux at distance $D$ approximately given by
\be
F_{\nu}(t) = \frac{2\pi h\nu^{3}}{c^{2}}\frac{\pi r_{\rm ph}^{2}}{D^{2}}\frac{1}{\exp[h\nu/kT_0(t)] - 1}
\ee
In detail the spectrum deviates moderately from a Planckian form due to the relativistic radial motion of the fluid (\citealt{Goodman86}, \citealt{Li&Sari08}, \citealt{Beloborodov11}).  

%\begin{figure}
%\includegraphics[width=0.5\textwidth]{tm_vs_a_plot.png}
%\centering
%\caption{Merger time as a function of binary separation a}
%\label{figure:merger_time}
%\end{figure}

%\begin{figure}
%\includegraphics[width=0.5\textwidth]{tlc_vs_a.png}
%\centering
%\caption{Ratio of expansion time of flow to photosphere to merger time as function of separation a}
%\label{figure:expansion_{\rm m}erger}
%\end{figure}

\section{A Short Burst of Gamma-Rays}
\label{sec:GRB}

The radiated luminosity is sharply peaked near the end of the inspiral on a timescale of milliseconds.  For $B_{\rm d} \sim 10^{13}-10^{14}$ G the emission reaches a peak luminosity of $L \simeq \dot{E}(a_{\rm min} \approx 30{\rm km}) \approx 10^{45}-10^{47}$ erg s$^{-1}$, at which time the spectral peak of $3k T_0 \sim 30-300$ keV is in the hard X-ray/soft gamma-ray band.  This duration and photon energy range is similar to those which characterize short GRBs, but the isotropic luminosities are $3-7$ orders of magnitude smaller.  Fig.~\ref{fig:contour} summarizes the emission power $\sim \dot{E}$ and average photon energy $3kT_0$ in the space of $B_{\rm d}-a$.

 The total bolometric fluence of the burst can be estimated as 
\begin{eqnarray}
F_{\rm tot} = \frac{\int_{a_{\rm min}}^{\infty}\dot{E}|\frac{dt}{da}|da}{4\pi D^{2}} \approx 3\times 10^{-9}\left(\frac{B_{\rm d}}{10^{14}\,{\rm G}}\right)^{2}\left(\frac{a_{\rm min}}{30{\rm km}}\right)^{-5/2}\left(\frac{D}{\rm 30\,Mpc}\right)^{-2}\,{\rm erg\,cm^{-2}},
\label{eq:fluencetot}
\end{eqnarray}
while that released external to the separation $a > a_{\rm min}$ is given by
\begin{eqnarray}
F(a) &\approx& 3\times 10^{-9}\left(\frac{B_{\rm d}}{10^{14}\,{\rm G}}\right)^{2}\left(\frac{a}{30{\rm km}}\right)^{-5/2}\left(\frac{D}{\rm 30\,Mpc}\right)^{-2}\,{\rm erg\,cm^{-2}},
\nonumber \\
 &\approx& 6\times 10^{-9}\left(\frac{B_{\rm d}}{10^{14}\,{\rm G}}\right)^{2}\left(\frac{t_{\rm m}}{\rm ms}\right)^{-5/8}\left(\frac{D}{\rm 30\,Mpc}\right)^{-2}\,{\rm erg\,cm^{-2}}.
\label{eq:fluence}
\end{eqnarray}

The duration of the burst measured by a gamma-ray detector is approximately the time that the flux remains above the detection threshold.  Following \citet{Band03}, we estimate that the photon detection sensitivity of BATSE and {\it Swift} BAT is approximately constant (to within a factor of $\lesssim 2$) across the energy range $\sim  100-1000$ keV of interest, at a value of $\sim 1(\Delta t/1{\rm s})^{-1/2}$ photons cm$^{-2}$ s$^{-1}$, where $\Delta t$ is the integration time.  Taking the relevant integration time at a given luminosity to be the inspiral time over which $\dot{E}$ changes, $\Delta t= t_{\rm m}$, we estimate a fluence sensitivity of (for $r_0 = a$)
\begin{eqnarray}
F_{\rm lim} &\approx& 1.0\times\left(\frac{t_{\rm m}}{1\,{\rm s}}\right)^{-1/2} \times  (3 k T_0) \times t_{\rm m}  \nonumber \\
&\approx& 3\times 10^{-8}\left(\frac{B_{d}}{10^{14}{\rm G}}\right)^{1/2}\left(\frac{t_{\rm m}}{\rm ms}\right)^{-1/32}\,{\rm erg\,cm^{-2}}
\label{eq:Flim}
\end{eqnarray}
This sensitivity is indeed comparable to the lowest measured GRB fluences (e.g., \citealt{vonKienlin+14}).

Equating~(\ref{eq:fluence}) and (\ref{eq:Flim}) gives the time interval that the burst is detectable above the instrument threshold, i.e. the burst duration,
\be
T_{\rm b} \approx 0.07\left(\frac{B_{\rm d}}{10^{14}\,{\rm G}}\right)^{48/19}\left(\frac{D}{\rm 30\,Mpc}\right)^{-64/19}\,{\rm ms}
\label{eq:correlation}
\ee
%Combining this with equation (\ref{eq:fluence}) and eliminating the distance results in a predicted duration-fluence relationship of
%\be
%F_{\rm tot} \approx 5\times 10^{-9}\left(\frac{T_{\rm b}}{\rm ms}\right)^{19/32}\,{\rm erg\,cm^{-2}},
%\ee
%which is independent of $B_{\rm d}$.

The burst is detected at all only if $T_{\rm b} \gtrsim t_{\rm m}(a_{\rm min} \approx 30$ km), a condition which translates into a maximum detection distance of
\be
D_{\rm max} \approx 10 \left(\frac{B_{\rm d}}{10^{14}\,{\rm G}}\right)^{3/4}\,{\rm Mpc}
\ee
For a chirp source near the edge of the Advanced LIGO volume at $D \approx 200$ Mpc, the precursor gamma-ray burst is only detectable for unrealistic values of $B_{\rm d} \gtrsim 10^{15}$ G. 

\section{Discussion}

\label{sec:conclusions}
Since the precursor emission is roughly isotropic, the all-sky detection rate is estimated to be
\be
\dot{N} \approx \frac{4\pi}{3}D_{\rm max}^{3}\mathcal{R} \approx 0.004\left(\frac{B_{\rm d}}{10^{14}\,{\rm G}}\right)^{9/4}\left(\frac{\mathcal{R}}{1000\,{\rm Gpc^{-3}\,yr^{-1}}}\right)\,{\rm yr^{-1}},
\label{eq:Ndot}
\ee
where $\mathcal{R} \sim 10-10^{4}$ Gpc$^{-3}$ yr$^{-1}$ is the uncertainty range in the local volumetric rate of NS-NS mergers (\citealt{Abadie+10} for a review).  Even for the most optimistic merger rates, we see that average field strengths of $B_{\rm d} \gtrsim 2\times 10^{14}$ G are required for a single detection over the 10 year lifetime of a gamma-ray mission like BATSE or {\it Swift}.  On the other hand, if even a small fraction of merging binaries possess strongly magnetized NSs, these outliers will dominate the observed population.  As mentioned earlier, although dipole fields this strong are rare among the pulsar population, the higher multi-pole moments$-$which are those most relevant to final stages of the inspiral where most of the fluence is accumulated$-$are not well constrained observationally.

\citet{Tanvir+05} measured a statistically significant correlation between the sky positions of short GRBs detected by BATSE and local cosmic structure as quantified by catalogs of nearby galaxies.  In order to reproduce the observed correlation, they concluded that between 10 and 25 percent of short GRBs must originate within the local universe, at distances of $\lesssim 70$ Mpc.  It is tempting to associate the sub-population identified by \citet{Tanvir+05} with the precursors of off-axis NS-NS mergers described here.
Given the $\sim$ 700 short GRBs detected over the 9 year lifetime of BATSE, this local class of bursts must occur roughly 10 times per year.  Such a high rate is compatible with equation (\ref{eq:Ndot}) if the merger rate is at the high end of present estimates, and if a sizable fraction of merging NS-NS binaries are characterized by magnetar-strength fields of $B_{\rm d} \gtrsim$ few $10^{14}$ G.  

Because the relativistic jet produced in the post-merger phase would in most cases be pointed away from our line of sight, such events would not be characterized by luminous X-ray or optical afterglows and hence would not have their host galaxies identified.  Being relatively nearby, however, the jet or the mildly relativistic merger ejecta could still produce a detectable off-axis radio afterglow, on timescales of weeks to months (e.g.~\citealt{Nakar&Piran11}).  This idea could be tested by performing radio follow-up of short GRBs without detected X-ray afterglows. 
% Future work can also address whether a sub-population of host-less short GRBs obey the duration-fluence correlation predicted by equation (\ref{eq:correlation}).

%The magnetospheric precursor model also predicts that the local population of GRB would preferentially have short durations, with durations of tens of milliseconds, due to the rapid rise in the luminosity approaching the point of merger. However, \citet{Tanvir+05} find a somewhat stronger correlation with local substructure when considering short bursts with longer $T_{90}$ durations.  

We have focused here on thermal emission from the pair fireball because its properties are largely determined by the physical size and power output of the system, and since the emission properties (a millisecond burst of 100-1000 keV photons) are in fact already the most optimal characteristics one could hope for given existing wide-field gamma-ray detectors.  However, general relativistic MHD numerical simulations of the magnetosphere interaction (e.g.~\citealt{Palenzuela+13a}) show that a sizable fraction of the electromagnetic power is carried away to large radii $r \gg a$ by a large-scale Poynting flux along open magnetic field lines.  Initially, the outflow will resemble the hot electromagnetic solutions described by \citet{Thompson&Gill14}.  Depending on the magnetization of the outflow following pair annihilation freeze-out, energy  stored in the magnetic field could act to accelerate the remaining pairs to ultra-high Lorentz factors, $\Gamma \gtrsim 10^{3}-10^{4}$.  Dissipation within this magnetized outflow, occurring well above the pair photosphere, could give rise to particle acceleration and non-thermal emission, for instance due to forced magnetic reconnection (e.g.~\citealt{Sironi&Spitkovsky14}) when magnetized shells collide following the sharply rising outflow power, $\dot{E} \propto t_{\rm m}^{-13/8}$ (eq.~\ref{eq:edotmax}).

Near the base of the outflow, the pair density (eq.~\ref{eq:npair}) is orders of magnitude too high to allow low frequency radio emission to escape.  However, one may speculate that above the photosphere the conditions within such a highly magnetized, ultra-relativistic outflow could give rise to the conditions necessary for a coherent burst of radio emission (\citealt{Hansen&Lyutikov01}, \citealt{Zhang14}, \citealt{Wang+16}), similar to the observed class of fast radio bursts (FRB; \citealt{Lorimer+07,Keane+12,Thornton+13,Spitler+14}).  The total fluence of precursor event is several orders of magnitude more than that required to explain FRBs, even assuming they occur cosmological distances (eq.~\ref{eq:fluencetot}).  Future work is required to check whether the outflow can indeed achieve a sufficiently high bulk Lorentz factor, $\Gamma \gtrsim 10^{3}-10^{4}$ to overcome constraints due to induced Compton and Raman scattering on the optical depth of the radio burst (\citealt{Lyubarsky08}).  Wide-field radio arrays, such as LOFAR \citep{VanHaarlem+13}, provide nearly continuous coverage of the northern hemisphere sky in the hundreds of MHz band, and hence could be used to search for an FRB in response to a short GRB (\citealt{Zhang14}).  Even an FRB produced during the final stages of the pre-merger inspiral could arrive at Earth delayed by seconds or longer from the time of the GRB trigger due to the subluminal propagation time associated with its high dispersion measure.

In addition to magnetosphere interaction, tidal resonant excitation of modes in the NS crust of provides an additional way to tap into the orbital energy of the merging binary \citep{Tsang+12}.  If driven to non-linear amplitudes, such modes shatter the crust, releasing $\sim 10^{46}-10^{47}$ ergs up to tens of seconds prior to merger, also potentially producing an observable flare \citep{Tsang+12,Tsang13}.  Given the estimated fluence threshold of $\sim 10^{-8}$ erg cm$^{-2}$, we estimate that if a significant fraction of the crust energy is transmitted to high energy radiation and a pair fireball as envisioned here, then such events could be detectable to a comparable distance of tens of Mpc, and hence could also contribute a local sub-population of short GRBs, or as precursors to short GRBs \citep{Troja+10}.  The tidal excitation model can be distinguished from the magnetospheric model by the significant delay predicted between an observed precursor and the termination of in the inspiral as measured from the GW signal.

%\begin{figure}
%\includegraphics[width=0.5\textwidth]{gamma_flux_50_{\rm m}pc.png}
%\centering
%\caption{Flux during merger in $\gamma$-ray spectrum for Distance = 50 Mpc}
%\label{figure:flux_50}
%\end{figure}

%\begin{figure}
%\includegraphics[width=0.5\textwidth]{gamma_flux_100_{\rm m}pc.png}
%\centering
%\caption{Flux during merger in $\gamma$-ray spectrum for Distance = 100 Mpc}
%\label{figure:flux_100}
%\end{figure}

%\begin{figure}
%\includegraphics[width=0.5\textwidth]{gamma_flux_200_{\rm m}pc.png}
%\centering
%\caption{Flux during merger in $\gamma$-ray spectrum for Distance = 200 Mpc}
%\label{figure:flux_200}
%\end{figure}

%\begin{figure}
%\includegraphics[width=0.5\textwidth]{Luminosity_during_{\rm m}erger_gamma_range_D_50.png}
%\centering
%\caption{Luminosity during time above L threshold for GBM telescope.}
%\label{figure:L_thresh}
%\end{figure}

%\begin{figure}
%\includegraphics[width=0.5\textwidth]{Fluence_D_50.png}
%\centering
%\caption{Fluence at GBM telescope for D = 50 Mpc.}
%\label{figure:Fluence_50}
%\end{figure}

%\begin{figure}
%\includegraphics[width=0.5\textwidth]{Fluence_D_100.png}
%\centering
%\caption{Fluence at GBM telescope for D = 100 Mpc.}
%\label{figure:Fluence_100}
%\end{figure}

%\begin{figure}
%\includegraphics[width=0.5\textwidth]{Fluence_D_200.png}
%\centering
%\caption{Fluence at GBM telescope for D = 200 Mpc.}
%\label{figure:Fluence_200}
%\end{figure}

\section*{Acknowledgments}

We thank Nat Butler, Dan D'Orazio, Janna Levin, and Tony Piro for helpful conversations.  BDM gratefully acknowledges support from NASA Fermi grant NNX14AQ68G, NSF grant AST-1410950, NASA ATP grant NNX16AB30G, and the Alfred P. Sloan Foundation.

%\bibliographystyle{mn2e}
%\bibliography{./}
%\bibliography{ms}

%\begin{thebibliography}{}
%\end{thebibliography}

\end{document}